**On the two-step hybrid design for augmenting randomized trials using real-world data**

Jiapeng Xu, MS[1]

Ruben P.A. van Eijk, MD, PhD[1,2]

Alicia Ellis, PhD[3]

Tianyu Pan, PhD[1]

Lorene M. Nelson, PhD[4]

Kit C.B. Roes, PhD[5]

Marc van Dijk, MSc[3]

Maria Sarno, MD, FFPM[3]

Leonard H. van den Berg, MD, PhD[2]

Lu Tian, PhD[1]

Ying Lu, PhD[1,4]

1. Department of Biomedical Data Science and Center for Innovative Study Design, Stanford University School of Medicine, California, CA, USA
2. Department of Neurology, UMC Utrecht Brain Centre, University Medical Centre Utrecht, Utrecht, The Netherlands
3. UCB Pharma, Brussels, Belgium
4. Department of Epidemiology and Population Health, Stanford University School of Medicine, California, CA, USA
5. Section Biostatistics, Department of Health Evidence, Radboud Medical Centre, Nijmegen, The Netherlands

Corresponding author: Ying Lu, PhD, ylu1@stanford.edu; Ruben van Eijk, MD, PhD, r.p.a.vaneijk-2@umcutrecht.nl

## Abstract

Hybrid clinical trials, that borrow real-world data (RWD), are gaining interest, especially for rare diseases. They assume RWD and randomized control arm be exchangeable, but violations can bias results, inflate type I error, or reduce power. A two-step hybrid design first tests exchangeability, reducing inappropriate borrowing but potentially inflating type I error (Yuan et al., 2019). We propose four methods to better control type I error. Approach 1 estimates the variance of test statistics, rejecting the null hypothesis based on large sample normal approximation. Approach 2 uses a numerical approach for exact critical value determination. Approach 3 splits type I error rates by equivalence test outcome. Approach 4 adjusts the critical value only when equivalence is established. Simulation studies using a hypothetical ALS scenario, evaluate type I error and power under various conditions, compared to the Bayesian power prior approach (Ibrahim et al., 2015). Our methods and the Bayesian power prior control type I error, whereas Yuan et al. (2019) increases it under exchangeability. If exchangeability doesn't hold, all methods fail to control type I error. Our methods show type I error inflation of 6%-8%, compared to 10% for Yuan et al. (2019) and 16% for the Bayesian power prior.

# 1 Introduction

A rare disease is defined as a disease with a prevalence of fewer than 200,000 people in the United States. The low prevalence introduces challenges for carrying out research and developing pharmacologic treatments for these diseases (Burton et al., 2021). The use of real-world data (RWD) offers new opportunities to enhance randomized clinical trials (RCTs), and is of particular relevance for rare diseases (Liu et al., 2022). To use RWD as an additional comparator to the treatment arm in a randomized design, however, it is necessary to assume exchangeability between patients who were randomized to control and those patients who originate from the RWD source. When this assumption is violated, borrowing RWD can introduce bias in the estimated treatment effect, thereby inflating type I error or adversely affecting statistical power. Several Bayesian methods are available to limit these adverse effects and improve inference, including dynamic borrowing methods based on Bayesian (Viele et al., 2013, Ibrahim et al., 2015).

To facilitate exchangeability, matching methods, including the use of propensity scores, reduce the baseline differences between two populations (Rosenbaum and Rubin, 1983). These methods do not, however, address the potential differences in outcomes occurring after baseline. Hence, it is necessary to determine additionally the equivalence of outcomes in order to establish exchangeability. A two-step hybrid clinical trial design proposed by Yuan et al (2019) combines participants in the RCT control arm with their propensity-matched external controls only when equivalence criteria for the outcomes are met. Yuan et al. (2019) further demonstrated that outcome exchangeability was necessary to avoid the introduction of bias. Bias could occur by falsely combining incompatible data sources under the assumption of no unmeasured confounders in outcomes. The value of this approach was recently illustrated for a case study in amyotrophic lateral sclerosis (ALS) (van Eijk et al., 2023); utilizing a hybrid approached improved precision and reduced the study duration compared to the original clinical trial.

While the two-step hybrid design reduces bias in borrowing, it uses two statistical tests which are conditioned on exchangeability, which leads to inflation of type I error (Yuan et al. (2019). The primary aim of this paper, therefore, is to propose alternative statistical tests for two-step hybrid designs to efficiently control for type I error and reduce bias in the estimated treatment effects. Second, we aim to compare the operating characteristics of our proposed tests with Yuan et al. (2019) and the Bayesian power prior approach (Ibrahim et al., 2015). Comparisons will be made by conducting an extensive simulation study based on a hypothetical trial for Amyotrophic Lateral Sclerosis (ALS) utilizing a disease registry as the RWD source.

The paper is structured as follows: Section 2 describes the statistical model of the two-step hybrid design; Section 3 introduces four new methods; Section 4 describe the simulation study design and results; Section 5 provides discussion and conclusions.

# 2 A two-step hybrid design for RWD augmentation

## 2.1 The Statistical Model

Consider a randomized trial in which patients are randomly assigned to either a treatment arm or a control arm. The primary objective is to determine if there is a mean difference in a continuous outcome variable between the two arms. The statistical hypotheses are the following:

$$H_0: \Delta = \mu_t - \mu_c = 0 \quad vs. \quad H_1: \Delta = \mu_t - \mu_c \neq 0 \qquad (1)$$

where $\mu_t$ and $\mu_c$ represent the mean responses for the treatment and control arms, respectively.

We are interested in the scenarios where an external dataset is available to augment the control arm of a clinical trial. We assume that the external data perfectly aligns with the baseline characteristics of the trial participants (i.e., no selection bias) and that an identical outcome variable can be observed under the control condition, which is the same as the control arm of the trial. Van Eijk et al. (2023) provided an example of using concurrent registry data to supplement a randomized phase II trial after propensity score matching. Yuan et al. (2019) also provided an example using data from historical control arms of a similar disease and the same control condition after matching baseline characteristics to augment a new randomized phase III trial.

Without loss of generalizability, let's assume $X|S = s \sim N(\mu_s, \sigma_s^2)$, for $s = t, c, r$. Here, $X$ represents the random outcome variable, $S$ represents the data source, "$t$" represents treatment arm; "$c$" represents control arm, and "$r$" represents external data, "μ" represents the mean, and "$\sigma^2$" represents the variance.

Let $n_t$, $n_c$, and $n_r$ be the sample sizes, and let $\bar{X}_t$, $\bar{X}_c$, and $\bar{X}_r$ denote the sample means for the treatment arm, control arm, and external data, respectively. The test statistics for the clinical trial without augment of external data is

$$Y_1 = \bar{X}_t - \bar{X}_c \sim N(\mu_t - \mu_c, \sigma_t^2/n_t + \sigma_c^2/n_c) = N(\Delta, \sigma_{Y_1}^2) \quad (2).$$

Similarly, we can compare the outcomes between the control arm and external data based on:

$$Y_2 = \bar{X}_r - \bar{X}_c \sim N(\mu_r - \mu_c, \sigma_r^2/n_r + \sigma_c^2/n_c) = N(\delta, \sigma_{Y_2}^2) \quad (3).$$

To augment the trial with external data, we want to compare the treatment arm with a weighted average of the observed sample means of the control and external data under the assumption that they are exchangeable, i.e., $\delta = 0$. The augmented test statistics is

$$Y_3\,(w) = \bar{X}_t - (1-w)\bar{X}_c - w\bar{X}_r = Y_1 - wY_2$$

The optimal weight that minimizes the variance of $Y_3\,(w)$ is the least square regression coefficient, i.e,

$$w^* = \frac{cov(Y_1, Y_2)}{\sigma_{Y_2}^2} = \frac{\frac{\sigma_c^2}{n_c}}{\frac{\sigma_r^2}{n_r} + \frac{\sigma_c^2}{n_c}}.$$

Thus,

$$Y_3 = Y_3\,(w^*) = Y_1 - w^* Y_2 \sim N\left(\Delta - w^*\delta, \sigma_{Y_1}^2(1-\rho^2)\right) \qquad (4).$$

where $\rho = \frac{cov(Y_1,Y_2)}{\sigma_{Y_1}\sigma_{Y_2}} = \frac{\sigma_c^2/n_c}{\sqrt{\sigma_t^2/n_t+\sigma_c^2/n_c}\sqrt{\sigma_r^2/n_r+\sigma_c^2/n_c}}$ is the correlation coefficient between $Y_1$ and $Y_2$.

Therefore, the variance of $Y_3$ is always less than the variance of $Y_1$. When $\delta = 0$, $Y_3$ is a more efficient test statistics for the null hypothesis (1).

## 2.2 A Two-step Hybrid Test

Whether $Y_3$ is an appropriate test statistic depends on the assumption that $\delta = 0$. Yuan et al (2019) proposed a two-step hybrid test that first examines the equivalence for the outcome between external RWD and trial participant data in the control arm. In such an equivalence test, the null hypothesis is $H_{0,EQ}$: $\delta \neq 0$. If the null hypothesis of non-equivalence is rejected, we can use $Y_3$ to test the null hypothesis (1). Otherwise, we use $Y_1$ to test the null hypothesis. This is considered a two-step hybrid test, also known as the test-then-pool approach (Viele et al., 2013).

The most frequently used equivalence test is the two one-sided tests (TOST) procedure (Schuirmann, 1987). In a TOST, a positive equivalence boundary (denoted by $\delta_{EQ}$) is pre-specified, which is the largest effect size of interest such that results falling within this range are deemed equivalent and exchangeable. The null and the alternative hypotheses for TOST are:

$$H_{0,EQ}: |\mu_r - \mu_c| \geq \delta_{EQ} \ \ vs\, H_{1,EQ}: |\mu_r - \mu_c| < \delta_{EQ} \tag{5}$$

Let $\alpha_{EQ}$ be the significant level for equivalence test, the null hypothesis is rejected if

$$-\delta_{EQ} + Z_{1-\alpha_{EQ}}\hat{\sigma}_{Y_2} < Y_2 < \delta_{EQ} - Z_{1-\alpha_{EQ}}\hat{\sigma}_{Y_2} \tag{6}$$

where $Z_{1-\alpha}$ is the $(1-\alpha)^{th}$ quantile of the standard normal distribution. Denoting $\theta = \delta_{EQ} - Z_{1-\alpha_{EQ}}\hat{\sigma}_{Y_2}$, $H_{0,ET}$ is rejectd if $|Y_2| < \theta$. Figure 1 shows the borrowing probabilities for different equivalence boundaries and significance levels. Yuan et al (2019) proposed a two-step procedure for test $H_0$: $\Delta = 0$ conditioning on whether $|Y_2| < \theta$. More specifically, we will reject $H_0$ based on following algorithm:

$$\begin{cases} when\ |Y_2| \geq \theta, & reject\ H_0\ if\ |Y_1| > Z_{1-\frac{\alpha}{2}}\hat{\sigma}_{Y_1} \\ when\ |Y_2| < \theta, & reject\ H_0\ if\ |Y_3| > Z_{1-\frac{\alpha}{2}}\hat{\sigma}_{Y_3} \end{cases} \tag{7}$$

Because the equivalence test for $H_{0,EQ}$ does not test for $H_0$ directly, Yuan et al. (2019) suggested that the proposed test procedure in (7) maintains type I error below $\alpha$. Reformatting test statistics in (7) to $Y = Y_1 1_{\{|Y_2| \geq \theta\}} + Y_3 1_{\{|Y_2| < \theta\}} = Y_1 - w^* Y_2 1_{\{|Y_2| < \theta\}}$, the p-value expressed in Equation (8) under a perfect condition of equivalence assumption when $\delta = 0$:

$$P\left(|Y| > Z_{1-\frac{\alpha}{2}}\left(\hat{\sigma}_{Y_1} 1_{\{|Y_2| \geq \theta\}} + \hat{\sigma}_{Y_3} 1_{\{|Y_2| < \theta\}}\right) \middle| \Delta = 0, \delta = 0\right)$$

$$= P\left(|Y| > Z_{1-\frac{\alpha}{2}}\left(\hat{\sigma}_{Y_1}\sqrt{\{1 - \hat{\rho}^2 1_{\{|Y_2| < \theta\}}\}}\right) \middle| \Delta = 0, \delta = 0\right) \tag{8}$$

The following Theorem 1 proves that Yuan et al. (2019) inflates type I error rate.

**Theorem 1**. $P\left(|Y| > Z_{1-\frac{\alpha}{2}}\left(\hat{\sigma}_{Y_1}\sqrt{\{1 - \hat{\rho}^2 1_{\{|Y_2| < \theta\}}\}}\right) \middle| \Delta = 0, \delta = 0\right) > \alpha.$

Proof of Theorem 1 is presented in Appendix A.1.

Li et al (2020) revisited the test-then-pool method and provided guidance on the selection of non-inferiority margin $\delta_{EQ}$, on the type I error, and statistical power.

# 3 Alternative 2-step tests

To ensure proper control of the Type I error rate, we propose four alternative two-step tests under the null hypothesis condition.

## 3.1 Approach 1: A large sample approximation

In this approach, although $Y$ no longer follows a normal distribution, we still estimate the variance of $Y$ and use a $Z$-test to reject the null hypothesis when $|Y| > Z_{1-\frac{\alpha}{2}}\hat{\sigma}_Y$ based on large sample approximation to the standard normal distribution. The variance $\hat{\sigma}_Y$ can be derived by plugging-in estimations of sample means and variances from Section 2, using the following formula:

$$\sigma_Y^2 = V(Y) = V(Y_1) + V\left(w^* Y_2 1_{\{|Y_2| < \theta\}}\right) - 2Cov\left(Y_1, w^* Y_2 1_{\{|Y_2| < \theta\}}\right)$$

$$= \sigma_{Y_1}^2 - w^{*2}\left(\mathrm{E}\left(1_{|Y_2|<\theta} Y_2^2\right) + \left[E\left(1_{|Y_2|<\theta} Y_2\right)\right]^2\right) + 2w^{*2}\delta E\left(1_{|Y_2|<\theta} Y_2\right) \quad (9)$$

Here, $\mathrm{E}\left(1_{|Y_2|<\theta} Y_2^2\right)$ and $\left[E\left(1_{|Y_2|<\theta} Y_2\right)\right]^2$ can be calculated through numerical integration with $\mathrm{E}\left(1_{|Y_2|<\theta} Y_2\right) = \int_{-\theta}^{\theta} \frac{y}{\sigma_{Y_2}} \phi\left(\frac{y-\delta}{\sigma_{Y_2}}\right) dy$ and $\mathrm{E}\left(1_{|Y_2|<\theta} Y_2^2\right) = \int_{-\theta}^{\theta} \frac{y^2}{\sigma_{Y_2}} \phi\left(\frac{y-\delta}{\sigma_{Y_2}}\right) dy$, and $\phi(y)$ as the standard normal density function.

Derivation of Equation (9) is given in Appendix A.2.

**Theorem 2.** When $\delta = 0$, $E(Y)$ is an unbiased estimator of treatment effect $\Delta$ and is a more efficient test statistics than $Y_1$. When $\delta \neq 0$, $E(Y)$ is a biased estimator of $\Delta$. However, the bias is bounded by a pre-specified constant:

$$|E(Y) - \Delta| \leq w^* \theta \alpha_{EQ} \tag{10}$$

Additionally, when $\delta \neq 0$, the type I error is also bounded:

$$P\left(\left|\frac{Y}{\hat{\sigma}_Y}\right| > z_{1-\frac{\alpha}{2}} \middle| \Delta = 0, \delta \neq 0\right) \leq \frac{\alpha}{2} + \Phi\left(z_{\frac{\alpha}{2}} + \frac{w\theta\alpha_{EQ}}{\hat{\sigma}_Y}\right), \tag{11}$$

where $\Phi(\mathrm{x})$ is the cumulative standard normal distribution.

Proof of Theorem 2 is in Appendix A.3.

### 3.2 Approach 2: Exact threshold method

Li et al. (2020) proposed this approach, where a numerical critical value $z^*$ is derived to reject the null hypothesis of $H_0$: $\Delta = 0$, when $\delta = 0$.

$$P\left(Y > z^*\left(\sigma_{Y_1} 1_{1_{|Y_2| \geq \theta}} + \sigma_{Y_3} 1_{|Y_2| < \theta}\right) \mid \Delta = 0, \delta = 0\right)$$

$$= P\left(Y_1/\sigma_{Y_1} > z^*, |Y_2/\sigma_{Y_2}| \geq \theta/\sigma_{Y_2} \mid \Delta = 0, \delta = 0\right)$$

$$+ P\left(Y_3/\sigma_{Y_3} > z^*, |Y_2/\sigma_{Y_2}| < \theta/\sigma_{Y_2} \mid \Delta = 0, \delta = 0\right) = \alpha/2 \tag{12}$$

This calibration method depends on the joint distribution between $Y_1$ and $Y_2$, and $Y_3$ under the null hypothesis, and the rejection value of the equivalence test. When $\Delta = \delta = 0$,

$\begin{pmatrix} \frac{Y_1}{\sigma_{Y_1}} \\ \frac{Y_2}{\sigma_{Y_2}} \end{pmatrix} \sim N\left(\begin{pmatrix} 0 \\ 0 \end{pmatrix}, \begin{bmatrix} 1 & \rho \\ \rho & 1 \end{bmatrix}\right)$, and $\begin{pmatrix} \frac{Y_3}{\sigma_{Y_3}} \\ \frac{Y_2}{\sigma_{Y_2}} \end{pmatrix} \sim N\left(\begin{pmatrix} 0 \\ 0 \end{pmatrix}, \begin{bmatrix} 1 & 0 \\ 0 & 1 \end{bmatrix}\right)$, $z^*$ can be derived numerically.

The type I error and power of Approach 2 has been discussed in Li et al (2020). As they demonstrated in their paper, when $\delta \neq 0$, type I error will not be controlled.

### 3.3 Approach 3: Splitting type I error conditioning on outcomes of the equivalence test

In this approach, we pre-specify Type I error splitting based on whether we accept or reject the null hypothesis of non-equivalent outcomes between the control arm and RWD. Let $\upsilon$ ($0 < \upsilon < 1$) be the proportion of type I error spent to non-equivalence between control arm and RWD, and $1 - \upsilon$ on the alternative condition. The two critical values, $z_1^*$ and $z_2^*$ can be derived from Equations (13) and (14). Here, $z_1^*$ is the critical value for the scenario when the equivalence test accepted the null hypothesis, and $z_2^*$ is the critical value when the equivalence test rejects the null hypothesis. Specifically, they can be selected to satisfy

$$P\left(\frac{Y_1}{\sigma_{Y_1}} > z_1^* \middle| |Y_2| \geq \theta, \Delta = 0, \delta = 0\right) = \frac{\nu\alpha}{2\beta_{EQ}} \tag{13}$$

and

$$P\left(\frac{Y_3}{\sigma_{Y_3}} > z_2^* \middle| |Y_2| < \theta, \Delta = 0, \delta = 0\right) = \frac{(1-\nu)\alpha}{2(1-\beta_{EQ})} \tag{14}$$

where $\alpha$ is the significance level of the primary test, and $\beta_{\mathrm{EQ}}$ is the type II error of the equivalence test. Note that $\beta_{\mathrm{EQ}}$ can be calculated as

$$\beta_{EQ} = P(|Y_2| \geq \theta | \delta = 0) = 1 - \int_{-\frac{\theta}{\sigma_{Y_2}}}^{\frac{\theta}{\sigma_{Y_2}}} \phi(y)dy.$$

Then, we have

$$P(Reject\ H_0 | \Delta = 0, \delta = 0)$$

$$= P\left(\frac{Y_1}{\sigma_{Y_1}} > z_1^*, |Y_2| \geq \theta |, \Delta = 0, \delta = 0\right) + P\left(\frac{Y_3}{\sigma_{Y_3}} > z_2^*, |Y_2| < \theta | \Delta = 0, \delta = 0\right)$$

$$= P\left(\frac{Y_1}{\sigma_{Y_1}} > z_1^* \middle| |Y_2| \geq \theta, \Delta = 0, \delta = 0\right) P(|Y_2| \geq \theta | \Delta = 0, \delta = 0)$$

$$+P\left(\frac{Y_3}{\sigma_{Y_3}} > z_2^* \middle| |Y_2| < \theta, \Delta = 0, \delta = 0\right) P(|Y_2| < \theta | \Delta = 0, \delta = 0)$$

$$= \frac{\nu\alpha}{2\beta_{EQ}}\beta_{EQ} + \frac{(1-\nu)\alpha}{2(1-\beta_{EQ})}(1-\beta_{EQ}) = \frac{\alpha}{2} \qquad (15)$$

Approach 2 is a special case of Approach 3 with a unique υ ensuring $z_1^* = z_2^*$. Another special case is selecting $\nu = \beta_{EQ}$, splitting the Type I error equally between two decisions of the equivalence test.

The advantage of Approach 3 over Approach 2 is the pre-specified splitting of type I error rate under two conditions. However, like Approach 2, Approach 3 does not guarantee Type I error control when $\delta \neq 0$.

Table 1 showed the critical values for various splitting thresholds $\nu$ under different experimental conditions.

(Table 1 is here)

### 3.4 Approach 4: Adjusting critical value when borrowing RWD

When Δ= $\delta = 0$, $\begin{pmatrix} \frac{Y_1}{\sigma_{Y_1}} \\ \frac{Y_2}{\sigma_{Y_2}} \end{pmatrix} \sim N\left(\begin{pmatrix} 0 \\ 0 \end{pmatrix}, \begin{bmatrix} 1 & \rho \\ \rho & 1 \end{bmatrix}\right)$, and $\begin{pmatrix} \frac{Y_3}{\sigma_{Y_3}} \\ \frac{Y_2}{\sigma_{Y_2}} \end{pmatrix} \sim N\left(\begin{pmatrix} 0 \\ 0 \end{pmatrix}, \begin{bmatrix} 1 & 0 \\ 0 & 1 \end{bmatrix}\right)$. According to Appendix A.1, the type I error inflation equals:

$$2\left(\frac{1}{2\pi}\int_0^{\rho}(1-z^2)^{-1/2}\exp\left[-\frac{c_1^2 + 2c_1c_2z + c_2^2}{2(1-z^2)}\right]dz \right.$$
$$\left. - \frac{1}{2\pi}\int_0^{\rho}(1-z^2)^{-1/2}\exp\left[-\frac{c_1^2 - 2c_1c_2z + c_2^2}{2(1-z^2)}\right]dz\right) > 0,$$

where $c_1 = z_{1-\alpha/2}$ and $c_2 = \theta$. The amount of type I error inflation can be calculated numerically. Then, we choose an adjusted cutoff value $z^* > z_{1-\alpha/2}$ by solving the following equation

$$P\left(z^*\sigma_{Y_3} > Y_3 > z_{1-\alpha/2}\,\sigma_{Y_3}, |Y_2| < \theta | \Delta = 0, \delta = 0\right)$$
$$= 2\left(\frac{1}{2\pi}\int_0^{\rho}(1-z^2)^{-1/2}\exp\left[-\frac{c_1^2 + 2c_1c_2z + c_2^2}{2(1-z^2)}\right]dz\right.$$
$$\left. -\frac{1}{2\pi}\int_0^{\rho}(1-z^2)^{-1/2}\exp\left[-\frac{c_1^2 - 2c_1c_2z + c_2^2}{2(1-z^2)}\right]dz\right) \quad (16)$$

to cancel the inflated type I error, where $c_1 = z_{1-\alpha/2}$ and $c_2 = \theta$. We reject the null hypothesis based on $\left|\frac{Y_3}{\sigma_{Y_3}}\right| > z^*$ when $|Y_2| < \theta$.

The Type I error is calibrated to the pre-specified significance level while maintaining the critical value when the null hypothesis of the equivalence test is not rejected and RWD is not pooled.

# 4. Simulation Studies

## 4.1 Comparison of Methods

In this section, we conduct simulation studies to evaluate the performance characteristics of our proposed approaches and compare them with previously proposed methods, including the method proposed by Yuan et al. (2019) and the Bayesian power prior method proposed by Ibrahim et al. (2015).

The Bayesian power prior method (Ibrahim et al., 2015) is a common approach to augment the control arm of the RCT trial data (denoted as $D$) with external RWD (denoted as $D_0$). The power prior method estimates an informative prior for the $D$ based on an initial non-informative prior of treatment effect $\pi_0(\mu)$ and a likelihood function of the parameter given $D_0$ that is raised to a power ($\alpha_0$) between 0 and 1. The resulting 'posterior' is used as an informative prior for $D$, which can be described as:

$$\pi(\mu|D_0,\alpha_0) \propto L(\mu|D_0)^{\alpha_0}\pi_0(\mu) \tag{17}$$

where $\alpha_0$ is set based on the heterogeneity of $D_0$ and $D$ (dynamic borrowing) and acts as a discounting parameter controlling the influence of $D_0$ on $L(\theta|D)$. In our case, if the external registry data is inconsistent with the control arm, $\alpha_0$ tends to be small and vice versa. We choose $\alpha_0$ that maximizes the marginal likelihood function, defined as:

$$m(\alpha_0) = \frac{\int L(\mu|\mu_c,\sigma_c^2)L(\mu|\mu_r,\sigma_r^2)\pi_0(\mu)\,d\mu}{\int L(\mu|\mu_r,\sigma_r^2)\pi_0(\mu)\,d\mu} = \frac{\int \phi\left(\frac{\mu-\mu_c}{\sigma_c}\right)\phi^{\alpha_0}\left(\frac{\mu-\mu_r}{\sigma_r}\right)\phi\left(\frac{\mu-\mu_0}{\sigma_0}\right)d\mu}{\int \phi^{\alpha_0}\left(\frac{\mu-\mu_r}{\sigma_r}\right)\phi\left(\frac{\mu-\mu_0}{\sigma_0}\right)d\mu} \tag{18}$$

Once, we get the $\alpha_0$ that maximizes the above marginal likelihood function, the posterior estimation of the pooled treatment effect is $\bar{X}_t - (1-\alpha_0 w^*)\bar{X}_c - \alpha_0 w^*\bar{X}_r = Y_1 - \alpha_0 w^* Y_2$. We reject the null hypothesis based on Z-test of the posterior distribution of the pooled treatment effect.

## 4.2 Simulation Settings

The simulation study was motivated by a hypothetical, 1:1 randomized, placebo-controlled clinical trial in ALS (van Eijk et al., 2023). The primary outcome was survival, defined as time to death from any cause. A synthetic RWD data set was generated with similar distribution properties as the Netherlands ALS registry used in van Eijk et al. (2023). The simulation study received a waiver from the University Medical Center Utrecht and Stanford University institutional review board.

In each iteration of our simulation study, we used a set of common eligibility criteria in ALS (van Eijk et al., 2023) to define eligible patients, i.e., those having a symptom duration ≤ 24 months, age at enrollment between 18 and 80 years, and a %predicted vital capacity ≥ 65%. An eligible patient was enrolled in the trial based on their binary probability of trial enrollment according to the logistic regression propensity model in van Eijk et al. (2023). Enrollment was stopped after randomly selecting 200 eligible patients from the registry as trial participants according to the propensity of enrollment specified in van Eijk et al. (2023). The patients were then randomized into an active treatment or placebo arm following a 1:1 ratio. The survival distribution of the control arm was derived from a Cox regression model with the Breslow baseline survival curve according to observed survival time and baseline covariates of eligible patients (Breslow, 1972). The estimated survival distribution was used to generate a random survival time for each patient in the trial. Covariates ($\vec{U}$) used in the Cox regression model included the site of symptom onset, El Escorial classification, %predicted vital capacity, age at enrollment, symptom duration, the revised ALS functional rating scale (ALSFRS-R) total score, and body mass index (BMI). A patient with covariates $\vec{U}_i$ in the control arm has the survival function

$$S_c\left(t;\ \vec{U}_i\right) = \left(S_0(t)\right)^{e^{\vec{a}'\vec{U}_i}} \tag{19}$$

where $S_0$(t) was the Breslow baseline survival function and $\vec{a}$ was the vector of regression coefficients from the Cox proportional hazards model for the eligible patients in the synthetic registry data.

The survival times for the treated patients were simulated based on a treatment effect specified by the hazards ratio (HR) $\theta_T$. Specifically, the survival time for patients in the treatment arm with covariates $\vec{U}_i$ followed the survival distribution with the survival function

$$S_T(t;\ \vec{U}_i) = \left(S_c(t;\ \vec{U}_i)\right)^{\theta_T}. \tag{20}$$

Patients who were eligible but not enrolled to the trial formed the external RWD population. To identify concurrent matched external controls from the registry, a logistic regression was performed to generate the simulation-specific propensity scoring equation for the 200 patients who were enrolled in the simulated trial. In our simulations, we selected two hundred external control patients from the remaining eligible patients in the registry who were not enrolled for trial, based on their simulation-specific propensity scores using nearest neighbor matches to the propensity scores of the trial patients. The number of matched patients can be more or less than trial participants depending on real applications. Here we selected an equal number to align with work done in van Eijk et al. (2023). The survival time for the RWD external controls was generated according to the pre-specified difference between the control arm and RWD measured by HR $\theta_R$. Thus, the survival time for patients in the RWD with covariates $\vec{U}_i$ has the survival function:

$$S_R(t;\ \vec{U}_i) = \left(S_c(t;\ \vec{U}_i)\right)^{\theta_R} \tag{21}$$

Since our primary endpoint was event-driven, the trial ended only when a pre-specified number of events was reached to ensure 80% statistical power for $\theta_T = 0.67$. We administratively censored the trial participants after the trial was completed. Additionally, the matched external control patients were also administratively censored according to the length of follow-up in the registry or the length of trial follow-up time, whichever occurred first. The simulation study repeated each simulation condition 10,000 times, which led to a 95% confidence interval for a 5% type-I error rate as (4.02%, 5.98%) and a 95% confidence interval for 80% power as (79.22%, 80.78%). Figure 2 provides a flowchart of the simulation process.

In the simulation study, the equivalence boundary $\delta$ for the control and RWD was set at 0.3 in $\log(\theta_R)$ (i.e., HR between 0.74 to 1.35) and significance level was chosen as 10%, at which level the borrowing probability being 70%-80% when the control arm and the RWD external controls was equivalent, considering the borrowing probability should not be too low when they were indeed identical. The significance level of the primary test was chosen as 5% (two-sided). The hybrid control design was studied under multiple scenarios where the true $\log(\theta_T)$ between the treatment arm and control arm was chosen to range between 0 (HR=1) and −0.4 (HR=0.67) and the true in $\log(\theta_R)$ between the control arm and external control group varied from −0.7 (HR=0.50) to 0.7 (HR=2.0) in steps of 0.1.

The following scenarios were included in our simulation studies

1. Under the perfect null condition: there was no survival difference between the treatment arm and control arm nor the control arm and external RWD, i.e., $\log(\theta_T)=\log(\theta_R) = 0$.
2. Under the null condition within equivalence between control and RWD: there was no survival difference between the randomized treatment arm and control arm. The log-HR between control arm and external RWD was not zero but smaller than the equivalence margin, i.e., $\log(\theta_T) = 0$ and $0<|\log(\theta_R)| < \delta$.
3. Under the null condition but no equivalence between control and RWD: there was no survival difference between the treatment arm and control arm. However, the control arm and external RWD had different survival distribution beyond equivalent margin, i.e., $\log(\theta_T) = 0$ and $|\log(\theta_R)| \geq \delta$.
4. Under the perfect alternative condition: there was a survival difference between the treatment arm and control arm but no survival difference between the control arm and external RWD, i.e., $\log(\theta_T) \neq 0$ and $\log(\theta_R) = 0$. The range of $\log(\theta_T)$ was -0.7 to 0.7, corresponding to $\theta_T$ from 0.50 to 2.01. The range was similar for Scenarios 5 and 6.

5. Under the alternative condition within equivalence between control and RWD: there was a survival difference between the treatment arm and control arm. However, the log-HR between control arm and external RWD was not zero but smaller than the equivalence merging, i.e., $\log(\theta_T) \neq 0$ and $|\log(\theta_R)| < \delta$.
6. Under the alternative condition but no equivalence between control and RWD: There was a survival difference between the treatment arm and control arm. Additionally, the control arm and external RWD had different survival beyond the equivalent margin, i.e., $\log(\theta_T) \neq 0$ and $|\log(\theta_R)|_c \geq \delta$.

We used the method described above to generate survival time for trial participants and external controls. We performed a Cox regression model using the control arm as the reference group. We included two binary group indicators, one for treatment patients and another one for the RWD controls as covariates in the Cox regression analysis. The estimated Cox regression coefficients for treatment arm and RWD were the corresponding $Y_1$ and $Y_2$, respectively in formulations in Section 2 and 3. The estimated covariance matrix corresponds to $\hat{\sigma}_{Y_1}^2, \hat{\sigma}_{Y_2}^2$, and $\hat{\rho}\hat{\sigma}_{Y_1}\hat{\sigma}_{Y_2}$, respectively. By plugging in these statistics, we calculated all the test statistics accordingly.
For the power prior described in Section 4.1, we used $Y_1 - Y_2$ and $Y_2$ as our reference. Here, $Y_1 - Y_2$ represented the log-HR between treatment arm and external control arm, and $Y_2$ represented the log-HR between control arm and external control arm.

### 4.3 Simulation Results

Figure 3 presents the probability of rejecting the inequivalence test and pooling using RWD to augment RCT data and the power prior $\alpha_0$, under various values of $\log(\theta_R)$. It showed that the maximum amount borrowing occurs when $\log(\theta_R)$=0. The chance decreased as $\log(\theta_R)$ moves away from 0. Even in the perfect case of $\log(\theta_R)$=0, the probability of test-then-pool approach had only 74.94% chance to borrow the RWD, which is the power of the equivalence test. The power prior was also below 1, reflecting the effect of chance. When $|\log(\theta_R)| = \delta = 0.3$, the test-then-pool approach had 10% of borrowing as specified type I error for the equivalence test. When $|\log(\theta_R)| \geq 0.5$, there was almost no borrowing by the 2-step hybrid approach. However, the power prior continued to borrow RWD beyond $|\log(\theta_R)| \geq 0.5$. One advantage of the 2-step hybrid approach is that it allows control of the amount of borrowing by changing the equivalence boundary to bound the potential bias.

(Figure 3 here)

Figure 4 displays the bias introduced for both 2-step hybrid approach and Bayesian power prior in two scenarios: when the exchangeability assumption held and when it was violated. It demonstrated that bias introduced in 2-step hybrid approach in treatment estimation was 0 when $\log(\theta_R) = 0$. As $\log(\theta_R)$ deviated from 0, the bias gradually increased, peaked around $|\log(\theta_R)| = 0.2$. Subsequently, the bias diminished due to reduced chance of borrowing. When $\log(\theta_R)$ was greater than 0, the bias introduced to underestimation of $\theta_T$ whereas when $\log(\theta_R)$ was less than 0, the bias led to overestimation of $\theta_T$. When using the Bayesian power prior, it was observed that no bias was introduced when the exchangeability assumption was held. However, as the value of $\log(\theta_R)$ deviated from 0, the bias introduced surpassed that of the 2-step hybrid approach, reaching its peak around $|\log(\theta_R)|$ = 0.3. Additionally, bias persisted when $|\log(\theta_R)| \geq 0.5$ due to continued borrowing.

(Figure 4 here)

The simulation results are summarized in Tables 2 and 3, showing the probability of rejecting the null hypothesis under the null and alternative conditions, respectively. These results are also presented graphically in Figures 4 and 5.

Under the perfect null condition, Yuan et al. (2019) had a type I error 6.58%, a 1.58% inflation from the targeted 5% and outside of the 95% confidence interval. Among all the methods, approach 1 was the most conservative, with a type I error of 3.82%, below the intended 5%. Approach 2, 3, and 4 were nearly identical and resulted in a type I error rate of 4.76%, 4.85%, and 5.07%, respectively. The power prior had an inflated type I error of 5.22%. As the true difference between the control arm and external RWD increased, the type I error inflates for all the methods (Figure 4). Approach 1 was the most conservative method that maintained type I error almost below 6% in all scenarios. The type I errors of approaches 2, 3, and 4, were similar and were at their maximum when log($\theta_R$)=0.2 with type I error rates of 8.13%, 7.93%, and 8.05%, respectively. Yuan's et al (2019) method had a type I error of 9.92%, whereas the power prior led to a serious type I error inflation of 16.78%. Overall, type I error inflation could not be fully controlled by any of the methods as the information borrowing may introduce bias into the treatment effect estimation. However, approach 1 could provide an upper bound of type I inflation, which could be pre-specified at trial design stage. As the true difference increased further, the type I error inflation started to decrease, which was due to a decreased likelihood of borrowing.

(Table 2 and Figure 5 is here)

Under the alternative condition when $\log(\theta_T)$ =-0.4, all methods led to improved power compared to a no borrowing scenario when $\log(\theta_R) = 0$. Among them, approach 1 had the least power gain of 82.73% and Bayesian power prior had the highest power of 92.51%. Approaches 2 to 4 had a statistical power of approximately 86%, which was 88% for when applying the Yuan et al method (Yuan et al., 2019).

When $\log(\theta_R) < 0$, all methods had decreased statistical power, which could even be below the designed trial power without borrowing. This was primarily due to improved survival of patients in RWD compared to patients in the trial, leading to an underestimation of treatment effects by pooling RWD.

The power gain was less for approaches 1-4 when $\log(\theta_R) > 0$ because of a rapid drop in probability of pooling RWD and the more stringently critical values to reject the null. Both Yuan et al. (2013) and Bayesian power prior (Ibrahim et al., 2015) had more power gain.

(Table 3 and Figure 6 is here)

## 5. Conclusion and Discussion

In clinical trials, borrowing RWD may be an attractive option as it may improve efficiency without increasing the number of trial participants, which is particularly beneficial for rare diseases. A critical assumption in borrowing RWD is exchangeability between the external controls and the control arm in the RCT. The two-step hybrid trial design includes an equivalence test for the exchangeability in outcomes. The equivalence test offers flexibility in choosing the equivalence margin, representing the maximum difference that is clinically acceptable for information borrowing. By selecting appropriate equivalence margin, one can control the chance of borrowing to balance the need of efficiency improvement and bias control. In practice, concurrent registries, such as the Netherlands ALS registry, could provide high-quality external RWD, as trial participants are a random sample from the registry in the same geographical area and temporal period, thus limiting potential sources of bias.

In this paper, we proposed four calibration methods for the test-then-pool approach to control type I error, conditioning on the results from the equivalence test. In addition, we conducted a simulation study using the two-step hybrid controlled clinical trial design proposed by Yuan et al. (2019) to investigate the operating characteristics under various scenarios. The simulation studies investigated the borrowing probability, type I error, and power under both exchangeable and non-exchangeable scenarios between the RCT control arm and RWD controls. Our simulation utilized synthetic data that was distributionally similar to the Netherlands ALS registry to closely represent a real-world setting.

Our simulation results showed that Yuan et al. (2019) resulted in inflated type I error even under exchangeable condition, and both Yuan et al. (2019) and Bayesian power prior approaches (Ibrahim et al., 2015) inflated type I error more than our proposed approaches under non-exchangeable condition, even though they were more likely to improve statistical power under the alternative hypothesis. The four proposed approaches in the Section 3 improved the power of the RCT while maintaining the type I error when the RCT control arm and RWD controls were exchangeable. Approach 1 was the most conservative method among the four, with a larger range of fully controlled non-exchangeable conditions, enabling it to control the type I error even if a small mean difference existed between the RCT control arm and the RWD controls. Conversely, approaches 2, 3, and 4 could only control the type I error if the mean difference was very close to 0. Therefore, if there is a strong belief that the RWD shares the same distribution as the RCT control arm, approaches 2-4 are preferred choices. However, if one wants to tightly control the type I error, it is recommended to use approach 1. In addition, when there is a large mean difference, approaches 1 and 4 can also ensure that the type I error remains commensurate with the pre-specified level and that there will be no loss in power. In contrast, approach 2 and 3 set the adjusted critical values conditioning on the result of the equivalence test, resulting in a lower type I error than the pre-specified level when significant mean differences existed and RWD data should not be pooled. Overall, approach 4 is a reasonable compromise in overall performance with a lower boundary of type I error inflation than the Yuan et al. (2019) and Bayesian power prior (Ibrahim et al., 2015) methods, yet achieves reasonable statistical power beyond approaches 1, 2, and 3 across a range of non-interchangeability scenarios.

One challenge in using the test-then-pool approach is the requirement for an adequate sample size in the RCT control arm to assure the power for the equivalence test, which may be difficult to obtain for rare diseases. In this scenario, Yuan et al. (2019) suggested using a two-stage design with an initially higher proportion patients to the treatment arm. If the interim analysis on equivalence test fails in rejecting the null, a reverse randomization ratio is proposed to ensure adequate power for the RCT. A more practical approach is to perform equal randomization in Stage 1 to assure that the number of patients in the RCT control arm is adequate for the equivalence test. If the control arm is exchangeable with the RWD controls, it is then safe to randomize more patients to the treatment arm.

In summary, using a two-step hybrid design that tests the exchangeability before pooling is a practical strategy and has shown the potential to improve the precision of the treatment effect estimate and the require study duration in an ALS trial (van Eijk et al., 2023). Appropriate statistical procedures are necessary to ensure the proper type I error rate of the test procedure. While we demonstrated four frequentist approaches, other approaches have been developed, including the Bayesian power prior, a popular method that does not require a test before pooling. Our simulation study showed the Bayesian power prior fails to control for type I error inflation even under ideal exchangeable condition. For drug development, it is critical to consider type I error controls for regulatory approval of any drugs. In real applications, sensitivity analysis using multiple methods should be performed to assure the robustness of findings instead relying on a single method.

## Software

The statistical codes underlying this publication were made open-source available on GitHub ("HybridDesign_CalibrationAndSimulation"; https://github.com/CISD-Stanford/HybridDesign_CalibrationAndSimulation).

### Funding Statement

This work was funded by the UCB-Stanford Digital Health Research Collaborative. The Sponsor of this study was not involved in the study design; in the collection and analysis; nor in the decision to submit the paper for publication. Scientific employees of the Sponsor contributed to the interpretation of study findings and writing the report.

### Competing interests

The author(s) has/have no competing interests to declare.

# Appendix

## A.1 Proof of Theorem 1

**Theorem 1**: $P\left(|Y| > Z_{1-\frac{\alpha}{2}}\left(\hat{\sigma}_{Y_1}\sqrt{\left\{1-\hat{\rho}^2 1_{\{|Y_2|<\theta\}}\right\}}\right)\middle|\Delta = 0, \delta = 0\right) > \alpha$ (A1.1)

***Proof.***

When $|Y_2| \geq \theta$, the left-hand side of the inequality equals

$$P\left(|Y_1| > Z_{1-\frac{\alpha}{2}}\hat{\sigma}_{Y_1}\middle||Y_2| \geq \theta, \Delta = 0, \delta = 0\right)$$

Similarly, when $|Y_2| < \theta$, we have

$$P\left(|Y_3| > Z_{1-\frac{\alpha}{2}}\hat{\sigma}_{Y_3}\middle||Y_2| < \theta, \Delta = 0, \delta = 0\right)$$

Therefore,

$$P\left(|Y| > Z_{1-\frac{\alpha}{2}}\left(\hat{\sigma}_{Y_1}\sqrt{1-\hat{\rho}^2 1_{\{|Y_2|<\theta\}}}\right)\middle|\Delta = 0, \delta = 0\right)$$

$$= P\left(|Y_1| > Z_{1-\frac{\alpha}{2}}\hat{\sigma}_{Y_1}, |Y_2| \geq \theta\middle|\Delta = 0, \delta = 0\right)$$

$$+ P\left(|Y_3| > Z_{1-\frac{\alpha}{2}}\hat{\sigma}_{Y_3}, |Y_2| < \theta\middle|\Delta = 0, \delta = 0\right)$$

$$= \alpha - P\left(|Y_1| > Z_{1-\frac{\alpha}{2}}\hat{\sigma}_{Y_1}, |Y_2| < \theta\middle|\Delta = 0, \delta = 0\right)$$

$$+ P\left(|Y_3| > Z_{1-\frac{\alpha}{2}}\hat{\sigma}_{Y_3}, |Y_2| < \theta\middle|\Delta = 0, \delta = 0\right)$$

Too prove the theorem, it is equivalent to prove

$$P\left(|Y_3| > Z_{1-\frac{\alpha}{2}}\hat{\sigma}_{Y_3}, |Y_2| < \theta\middle|\Delta = 0, \delta = 0\right) > P\left(|Y_1| > Z_{1-\frac{\alpha}{2}}\hat{\sigma}_{Y_1}, |Y_2| < \theta\middle|\Delta = 0, \delta = 0\right)$$

(A. 1.2)

By standardizing the normal random variables, it gives us

$$P\left(|Z_3| > Z_{1-\frac{\alpha}{2}}, |Z_2| < \theta\right) > P\left(|Z_1| > Z_{1-\frac{\alpha}{2}}, |Z_2| < \theta\right), \quad \text{(A.1.3)}$$

where $Z_1 = \frac{Y_1}{\widehat{\sigma_{Y_1}}}$, $Z_2 = \frac{Y_2}{\widehat{\sigma_{Y_2}}}$, $Z_3 = \frac{Y_3}{\widehat{\sigma_{Y_3}}}$, and they follow standard normal distribution. In addition, $Z_1$ *and* $Z_2$ are correlated with correlation coefficient $\rho$, and $Z_3$ *and* $Z_2$ are independent.

The left-hand side of the inequality (A.1.3) equals to

$$P\left(|Z_3| > Z_{1-\frac{\alpha}{2}}, |Z_2| < \theta\right) = P\left(|Z_3| > Z_{1-\frac{\alpha}{2}}\right) P(|Z_2| < \theta) = \alpha\big(\Phi(\theta) - \Phi(-\theta)\big) \quad (A.1.4)$$

The right-hand side of the inequality (A.1.3) equals to

$$P\left(|Z_3| > Z_{1-\frac{\alpha}{2}}, |Z_2| < \theta\right) = 2\left(P\left(Z_3 < -Z_{1-\frac{\alpha}{2}}, Z_2 < \theta\right) - P\left(Z_3 < -Z_{1-\frac{\alpha}{2}}, Z_2 < -\theta\right)\right)$$

$$= 2\left(\Phi\left(-Z_{1-\frac{\alpha}{2}}, \theta, \rho\right) - \Phi\left(-Z_{1-\frac{\alpha}{2}}, -\theta, \rho\right)\right) \quad (A.1.5)$$

According to Owen (1956), $\Phi(h,k,\rho)$ can be reduced to a single integral:

$$\Phi(h,k,\rho) = \frac{1}{2\pi}\int_0^{\rho} (1-z^2)^{-\frac{1}{2}} e^{-\frac{h^2-2hkz+k^2}{2(1-z^2)}}\, dz + \Phi(h)\Phi(k)$$

Hence, Equation (A.1.5) equals to

$$P\left(|Z_3| > Z_{1-\frac{\alpha}{2}}, |Z_2| < \theta\right)$$

$$= 2\left(\left(\frac{1}{2\pi}\int_0^{\rho}(1-z^2)^{-1/2}e^{-\frac{c_1^2+2c_1c_2z+c_2^2}{2(1-z^2)}}dz + \Phi(-c_1)\Phi(c_2)\right)\right.$$

$$\left.-\left(\frac{1}{2\pi}\int_0^{\rho}(1-z^2)^{-1/2}\mathrm{e}^{--\frac{c_1^2-2c_1c_2z+c_2^2}{2(1-z^2)}}dz + \Phi(-c_1)\Phi(-c_2)\right)\right)$$

$$= 2\left(\frac{1}{2\pi}\int_0^{\rho}(1-z^2)^{-1/2}\exp\left[-\frac{c_1^2+2c_1c_2z+c_2^2}{2(1-z^2)}\right]dz\right.$$

$$\left.-\frac{1}{2\pi}\int_0^{\rho}(1-z^2)^{-1/2}\exp\left[-\frac{c_1^2-2c_1c_2z+c_2^2}{2(1-z^2)}\right]dz\right)$$

$$+ 2\,\Phi(-c_1)\big(\Phi(c_2)-\Phi(-c_2)\big),$$

where $c_1 = Z_{1-\frac{\alpha}{2}}$ and $c_2 = \theta$

According to Equation (A.1.4), we know

$$2\,\Phi(-c_1)\big(\Phi(c_2)-\Phi(-c_2)\big) = \alpha\big(\Phi(\theta)-\Phi(-\theta)\big) = P\left(|Z_3| > Z_{1-\frac{\alpha}{2}}, |Z_2| < \theta\right)$$

That is, to prove Equation (A.1.3), we need to prove

$$\frac{1}{2\pi}\int_0^{\rho}(1-z^2)^{-1/2}\exp\left[-\frac{c_1^2-2c_1c_2z+c_2^2}{2(1-z^2)}\right]dz)$$

$$> \frac{1}{2\pi}\int_0^{\rho}(1-z^2)^{-1/2}\exp\left[-\frac{c_1^2+2c_1c_2z+c_2^2}{2(1-z^2)}\right]dz$$

Define the integrands $f(z)$ and $g(z)$:

$$f(z) = (1-z^2)^{-1/2}e^{-\frac{c_1^2-2c_1c_2z+c_2^2}{2(1-z^2)}}$$

$$g(z) = (1-z^2)^{-1/2}e^{-\frac{c_1^2+2c_1c_2z+c_2^2}{2(1-z^2)}}$$

Then, we have

$$\frac{f(z)}{g(z)} = \frac{(1-z^2)^{-1/2} e^{-\frac{c_1^2-2c_1c_2z+c_2^2}{2(1-z^2)}}}{(1-z^2)^{-1/2} e^{-\frac{c_1^2+2c_1c_2z+c_2^2}{2(1-z^2)}}} = e^{-\frac{c_1^2-2c_1c_2z+c_2^2}{2(1-z^2)}+\frac{c_1^2+2c_1c_2z+c_2^2}{2(1-z^2)}} = e^{\frac{2c_1c_2z}{1-z^2}}$$

Since $c_1 = Z_{1-\frac{\alpha}{2}}$ , $c_2 = \theta$ , $0 \leq z \leq \rho$ are all positive number, it follows that $e^{\frac{2c_1c_2z}{1-z^2}} > 1$.

Therefore,

$$f(z) > g(z)$$

And integrating both sides from 0 to $\rho$:

$$\int_0^{\rho} f(z)dz > \int_0^{\rho} g(z)dz$$

This is equivalent to:

$$\frac{1}{2\pi}\int_0^{\rho}(1-z^2)^{-1/2}\exp\left[-\frac{c_1^2-2c_1c_2z+c_2^2}{2(1-z^2)}\right]dz)$$

$$> \frac{1}{2\pi}\int_0^{\rho}(1-z^2)^{-1/2}\exp\left[-\frac{c_1^2+2c_1c_2z+c_2^2}{2(1-z^2)}\right]dz$$

Hence, we proved

$$P\left(|Y_3| > Z_{1-\frac{\alpha}{2}}\hat{\sigma}_{Y_3}, |Y_2| < \theta \middle| \Delta = 0, \delta = 0\right) > P\left(|Y_1| > Z_{1-\frac{\alpha}{2}}\hat{\sigma}_{Y_1}, |Y_2| < \theta \middle| \Delta = 0, \delta = 0\right)$$

This is equivalent to show

$$P\left(|Y| > Z_{1-\frac{\alpha}{2}}\left(\hat{\sigma}_{Y_1}\sqrt{1-\hat{\rho}^2 1_{\{|Y_2|<\theta\}}}\right) \middle| \Delta = 0, \delta = 0\right) > \alpha$$

**A.2 Derivation of variance of *Y***

The overall test statistics $Y$ equals:

$$Y = Y_1 - w^* Y_2 1_{\{|Y_2|<\theta\}}$$

Therefore, the variance of Y is:

$$Var(Y) = \sigma_{Y_1}^2 + w^{*2} V\left(Y_2 1_{\{|Y_2|<\theta\}}\right) - 2w^* Cov\left(Y_1, Y_2 1_{\{|Y_2|<\theta\}}\right)$$

$$= \sigma_{Y_1}^2 + w^2 \left(E\left[Y_2^2 1_{\{|Y_2|<\theta\}}\right] - E\left[Y_2 1_{\{|Y_2|<\theta\}}\right]^2\right)$$

$$- 2w\left(E\left[Y_1 Y_2 1_{\{|Y_2|<\theta\}}\right] - E[Y_1] E\left[Y_2 1_{\{|Y_2|<\theta\}}\right]\right)$$

Since

$$E\left[Y_1 Y_2 1_{\{|Y_2|<\theta\}}\right] = \boldsymbol{E}\left[\boldsymbol{E}\left[Y_1 Y_2 1_{\{|Y_2|<\theta\}} \middle| \boldsymbol{Y_2}\right]\right] = \boldsymbol{E}\left[Y_2 1_{\{|Y_2|<\theta\}} \boldsymbol{E}[Y_1 | \boldsymbol{Y_2}]\right]$$

$$= \boldsymbol{E}\left[Y_2 1_{\{|Y_2|<\theta\}} \left(\boldsymbol{\Delta} + \frac{\boldsymbol{cov(Y_1, Y_2)}}{\boldsymbol{\sigma_{Y_2}^2}} (\boldsymbol{Y_2} - \boldsymbol{\delta})\right)\right]$$

$$= (\boldsymbol{\Delta} - w^* \boldsymbol{\delta}) \boldsymbol{E}\left[Y_2 1_{\{|Y_2|<\theta\}}\right] + w^* E\left[Y_2^2 1_{\{|Y_2|<\theta\}}\right]$$

That is,

$$Var(Y) = \sigma_{Y_1}^2 + w^{*2} \left(E\left[Y_2^2 1_{\{|Y_2|<\theta\}}\right] - E\left[Y_2 1_{\{|Y_2|<\theta\}}\right]^2\right)$$

$$- 2w^{*2}\left(E\left[Y_2^2 1_{\{|Y_2|<\theta\}}\right] - \delta \boldsymbol{E}\left[Y_2 1_{\{|Y_2|<\theta\}}\right]\right) =$$

$$= \sigma_{Y_1}^2 - w^{*2}\left(\mathrm{E}[Y_2^2 1_{|Y_2|<\theta}] + E[Y_2 1_{|Y_2|<\theta}]^2\right) + 2w^{*2} \delta E[Y_2 1_{|Y_2|<\theta}$$

### A.3 Proof of Theorem 2

**Theorem 2.** When $\delta = 0$, $E(Y)$ is an unbiased estimator of treatment effect $\Delta$ and a more efficient test statistics than $Y_1$. When $\delta \neq 0$, $E(Y)$ is a biased estimator of $\Delta$. However, the bias is bounded by a pre-specified constant, i.e.,

$$|E(Y) - \Delta| \leq w^*\theta\alpha_{EQ}$$

In addition, when $\delta \neq 0$, the type I error bias is also bounded,

$$P\left(\left|\frac{Y}{\hat{\sigma}_Y}\right| > z_{1-\frac{\alpha}{2}} \middle| \Delta = 0, \delta \neq 0\right) \leq \Phi\left(z_{\frac{\alpha}{2}} + \frac{w\alpha_2\theta}{\hat{\sigma}_Y}\right) - \frac{\alpha}{2}$$

, where $\Phi(x)$ is the cumulative standard normal distribution.

***Proof.***

When $\delta = 0$, $E(Y_2) \ = \ 0 \ and \ E(Y) = \Delta - wP(|Y_2| < \theta)E\big(Y_2\big||Y_2| < \theta\big)$

Since $Z_2\big||Z_2| < \theta$ is symmetric around 0, $E\big(Y_2\big||Y_2| < \theta\big) = 0$

Therefore, $E(Y) = \Delta$ and $E(Y)$ is an unbiased estimator of treatment effect $\Delta$

When $\delta \neq 0$, the bias equals to

$$|E(Y) - \Delta| = \left|wE\big(Y_2 1_{\{|Y_2|<\theta\}}|\delta \neq 0\big)\right|$$

$$= \left|wE\big(Y_2\big||Y_2| < \theta, \delta \neq 0\big)P(|Y_2| < \theta|\delta \neq 0)\right|$$

$$\leq w\theta\alpha_{EQ}$$

For Approach 1, the type I error equals to

$$P\left(\left|\frac{Y}{\hat{\sigma}_Y}\right| > z_{1-\frac{\alpha}{2}}\middle|\Delta = 0, \delta \neq 0\right) = P\left(\frac{Y}{\hat{\sigma}_Y} > z_{1-\frac{\alpha}{2}}\middle|\Delta = 0, \delta \neq 0\right) +$$

$$P\left(\frac{Y}{\hat{\sigma}_Y} < -z_{1-\frac{\alpha}{2}}\middle|\Delta = 0, \delta \neq 0\right) = P\left(\frac{Y-E(Y)}{\hat{\sigma}_Y} > z_{1-\frac{\alpha}{2}} - \frac{E(Y)}{\hat{\sigma}_Y}\middle|\Delta = 0, \delta \neq 0\right) +$$

$$P\left(\frac{Y-E(Y)}{\hat{\sigma}_Y} < -z_{1-\frac{\alpha}{2}} - \frac{E(Y)}{\hat{\sigma}_Y}\middle|\Delta = 0, \delta \neq 0\right) = P\left(Z > z_{1-\frac{\alpha}{2}} - \frac{E(Y)}{\hat{\sigma}_Y}\middle|\Delta = 0, \delta \neq 0\right) +$$

$$P\left(Z < -z_{1-\frac{\alpha}{2}} - \frac{E(Y)}{\hat{\sigma}_Y}\middle|\Delta = 0, \delta \neq 0\right) = \Phi\left(z_{\frac{\alpha}{2}} + \frac{E(Y)}{\hat{\sigma}_Y}\right) + \Phi\left(z_{\frac{\alpha}{2}} - \frac{E(Y)}{\hat{\sigma}_Y}\right)$$

Since when $\Delta = 0, \delta \neq 0$, $|E(Y)|$ is bounded by $w\theta\alpha_{EQ}$, the type I error is also bounded by

$$P\left(\left|\frac{Y}{\hat{\sigma}_Y}\right| > z_{1-\frac{\alpha}{2}} \middle| \Delta = 0, \delta \neq 0\right) \leq \Phi\left(z_{\frac{\alpha}{2}} - \frac{w\theta\alpha_{EQ}}{\hat{\sigma}_Y}\right) + \Phi\left(z_{\frac{\alpha}{2}} + \frac{w\theta\alpha_{EQ}}{\hat{\sigma}_Y}\right) \leq \frac{\alpha}{2} + \Phi\left(z_{\frac{\alpha}{2}} + \frac{w\theta\alpha_{EQ}}{\hat{\sigma}_Y}\right)$$

Therefore, the type I error bias is also bounded by

$$\Phi\left(z_{\frac{\alpha}{2}} - \frac{w\theta\alpha_{EQ}}{\hat{\sigma}_Y}\right) + \Phi\left(z_{\frac{\alpha}{2}} + \frac{w\theta\alpha_{EQ}}{\hat{\sigma}_Y}\right) - \alpha$$

## Tables

**Table 1.** Type I error, adjusted α*, power, and borrowing probability for Approach 3 ($\sigma_t^2 = \sigma_c^2 = \sigma_r^2 = 1, n_t = 100, n_c = 100, n_r = 200$)

| | Type I error | Split proportion: 25% | | Split proportion: 50% | | Split proportion: 75% | | Borrowing Probability |
|---|---|---|---|---|---|---|---|---|
| | | Adjusted $\alpha^*$ | Power using $\alpha^*$ | Adjusted $\alpha^*$ | Power using $\alpha^*$ | Adjusted $\alpha^*$ | Power using $\alpha^*$ | |
| | | | | Equivalence Boundary $\delta =$ 0.25 | | | | |
| $\alpha_{EQ} = 0.05$ | 0.0599 | 0.0467 (0.0134/0.1216) | 0.7258 | 0.0440 (0.0274/0.0812) | 0.7827 | 0.0414 (0.0418/0.0406) | 0.8070 | 0.3082 |
| $\alpha_{EQ} = 0.10$ | 0.0658 | 0.0441 (0.0148/0.0678) | 0.7972 | 0.0390 (0.0314/0.0452) | 0.8229 | 0.0344 (0.0490/0.0226) | 0.8194 | 0.5526 |
| $\alpha_{EQ} = 0.15$ | 0.0673 | 0.0428 (0.0166/0.0548) | 0.8331 | 0.0363 (0.0362/0.0364) | 0.8422 | 0.0307 (0.0580/0.0182) | 0.8232 | 0.6850 |
| $\alpha_{EQ} = 0.20$ | 0.0670 | 0.0419 (0.0188/0.0488) | 0.8553 | 0.0347 (0.0426/0.0324) | 0.8542 | 0.0286 (0.0700/0.0162) | 0.8253 | 0.7697 |
| | | | | Equivalence Boundary $\delta =$ 0.30 | | | | |
| $\alpha_{EQ} = 0.05$ | 0.0663 | 0.0438 (0.0150/0.0648) | 0.8044 | 0.0385 (0.0320/0.0432) | 0.8268 | 0.0336 (0.0502/0.0216) | 0.8202 | 0.5790 |
| $\alpha_{EQ} = 0.10$ | 0.0672 | 0.0420 (0.0184/0.0496) | 0.8520 | 0.0350 (0.0414/0.0330) | 0.8525 | 0.0290 (0.0676/0.0166) | 0.8250 | 0.7572 |
| $\alpha_{EQ} = 0.15$ | 0.0656 | 0.0412 (0.0228/0.0446) | 0.8740 | 0.0335 (0.0542/0.0296) | 0.8646 | 0.0271 (0.0930/0.0148) | 0.8274 | 0.8424 |
| $\alpha_{EQ} = 0.20$ | 0.0635 | 0.0406 (0.0290/0.0420) | 0.8867 | 0.0329 (0.0734/0.0280) | 0.8719 | 0.0269 (0.1332/0.0140) | 0.8291 | 0.8921 |

**Table 2.** Simulation results under the null hypothesis: Type I error (in percentages) under different log HR between control arm and external control arm

| Log HR (trt/ctrl) | Log HR (ctrl/RWD) | No borrowing | Yuan et al. (2019) | Power Prior (2015) | Approach 1[a] | Approach 2[a] | Approach 3[a] | Approach 4[a] |
|---|---|---|---|---|---|---|---|---|
| 0 | 0.7 | 5.23 | 5.23 | 8.55 | 5.19 | 3.76 | 4.39 | 5.23 |
| 0 | 0.6 | 5.02 | 5.02 | 9.14 | 4.86 | 3.53 | 4.14 | 5.02 |
| 0 | 0.5 | 4.97 | 5.03 | 11.41 | 4.67 | 3.66 | 4.16 | 5.02 |
| 0 | 0.4 | 4.84 | 5.64 | 14.84 | 4.97 | 4.35 | 4.70 | 5.48 |
| 0 | 0.3 | 4.98 | 8.08 | 16.78 | 5.85 | 6.48 | 6.58 | 7.11 |
| 0 | 0.2 | 4.91 | 9.92 | 14.07 | 6.42 | 8.13 | 7.93 | 8.05 |
| 0 | 0.1 | 5.13 | 7.84 | 7.83 | 4.67 | 5.96 | 5.82 | 5.87 |
| **0** | **0** | **4.96** | **6.58** | **5.22** | **3.82** | **4.76** | **4.85** | **5.07** |
| 0 | -0.1 | 5.14 | 8.82 | 9.20 | 5.04 | 6.65 | 6.63 | 6.53 |
| 0 | -0.2 | 5.37 | 9.79 | 14.56 | 6.49 | 7.62 | 7.64 | 7.87 |
| 0 | -0.3 | 5.03 | 7.44 | 16.26 | 5.68 | 5.94 | 6.03 | 6.61 |
| 0 | -0.4 | 4.94 | 5.37 | 13.23 | 4.75 | 4.13 | 4.53 | 5.27 |
| 0 | -0.5 | 4.85 | 4.89 | 10.34 | 4.69 | 3.68 | 4.12 | 4.89 |
| 0 | -0.6 | 4.89 | 4.90 | 8.07 | 4.78 | 3.57 | 4.07 | 4.90 |
| 0 | -0.7 | 5.07 | 5.07 | 6.99 | 5.01 | 3.72 | 4.02 | 5.07 |

a. Approach 1: large sample approximate; Approach 2: exact threshold; Approach 3: splitting type I error rates; Approach 4: adjusting critical value when borrowing RWD.

**Table 3.** Simulation results under the alternative hypothesis: Power (in percentages) under different log HR between control arm and external control arm

| Log HR (trt/ctrl) | Log HR (ctrl/RWD) | No borrowing | Yuan et al. (2019) | Power Prior | Approach 1[a] | Approach 2[a] | Approach 3[a] | Approach 4[a] |
|---|---|---|---|---|---|---|---|---|
| −0.4 | 0.7 | 80.98 | 80.98 | 88.76 | 80.98 | 77.14 | 78.9 | 80.98 |
| −0.4 | 0.6 | 80.86 | 80.86 | 89.33 | 80.86 | 77.04 | 78.7 | 80.86 |
| −0.4 | 0.5 | 81.14 | 81.14 | 90.77 | 80.95 | 77.12 | 78.79 | 81.14 |
| −0.4 | 0.4 | 81.12 | 81.13 | 92.55 | 80.56 | 77.24 | 78.84 | 81.13 |
| −0.4 | 0.3 | 80.77 | 80.9 | 94.02 | 79.46 | 77.26 | 78.97 | 80.9 |
| −0.4 | 0.2 | 81.47 | 82.84 | 95.90 | 79.97 | 79.1 | 80.75 | 82.69 |
| −0.4 | 0.1 | 80.9 | 85.6 | 96.04 | 82.33 | 83.07 | 83.89 | 84.81 |
| **−0.4** | **0** | **80.74** | **88.43** | **92.51** | **82.73** | **86.39** | **86.05** | **85.62** |
| −0.4 | -0.1 | 80.47 | 86.12 | 84.56 | 77.74 | 82.99 | 82.5 | 81.7 |
| −0.4 | -0.2 | 80.41 | 81.29 | 73.87 | 75.05 | 77.8 | 78.02 | 78.37 |
| −0.4 | -0.3 | 80.96 | 80.73 | 69.00 | 77.15 | 76.76 | 77.78 | 79.89 |
| −0.4 | -0.4 | 80.79 | 80.78 | 69.61 | 78.95 | 76.72 | 77.94 | 80.64 |
| −0.4 | -0.5 | 80.63 | 80.63 | 70.79 | 79.83 | 76.65 | 77.89 | 80.63 |
| −0.4 | -0.6 | 80.53 | 80.53 | 72.33 | 80.35 | 76.85 | 77.71 | 80.53 |
| −0.4 | -0.7 | 80.58 | 80.58 | 73.30 | 80.54 | 76.72 | 77.49 | 80.58 |

a. Approach 1: large sample approximate; Approach 2: exact threshold; Approach 3: splitting type I error rates; Approach 4: adjusting critical value when borrowing RWD.

## Figures

**Figure 1.** Borrowing probability under different equivalence boundaries and significance levels

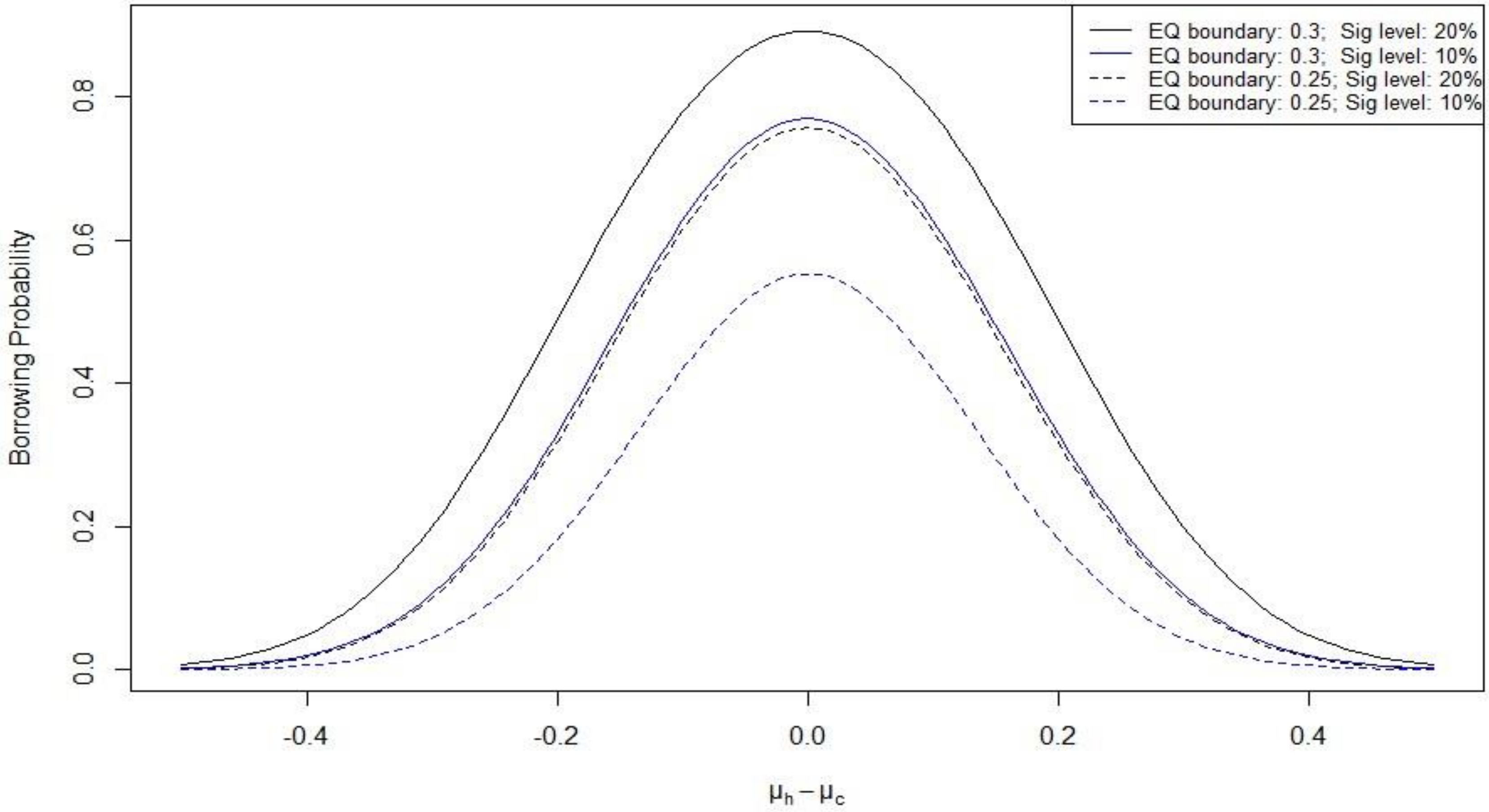

**Figure 2: Flow Chart of Simulation Data Generation**

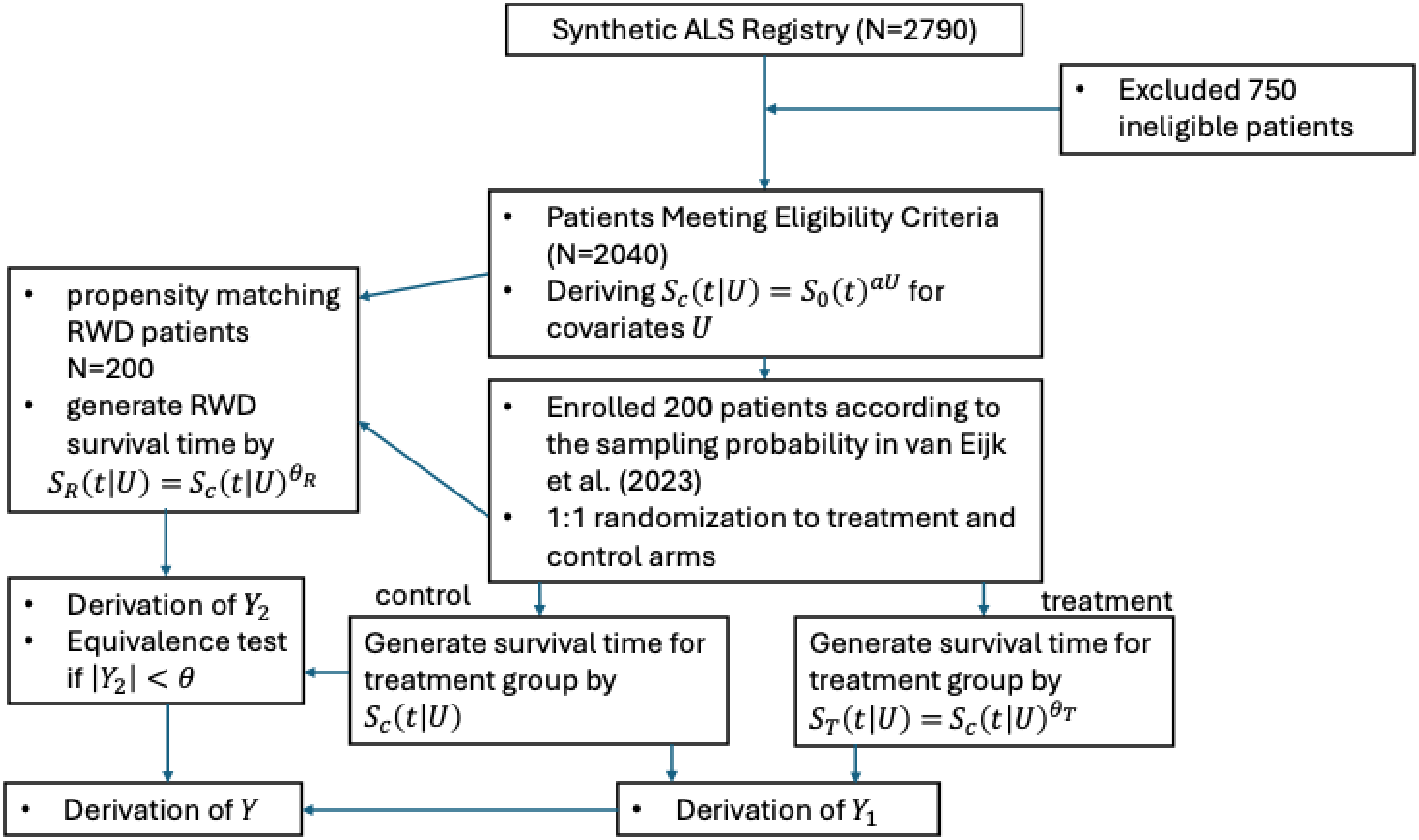

**Figure 3.** Borrowing probabilities and mean alpha (the discounting factor in power prior) under different log HR between control arm and external control arm

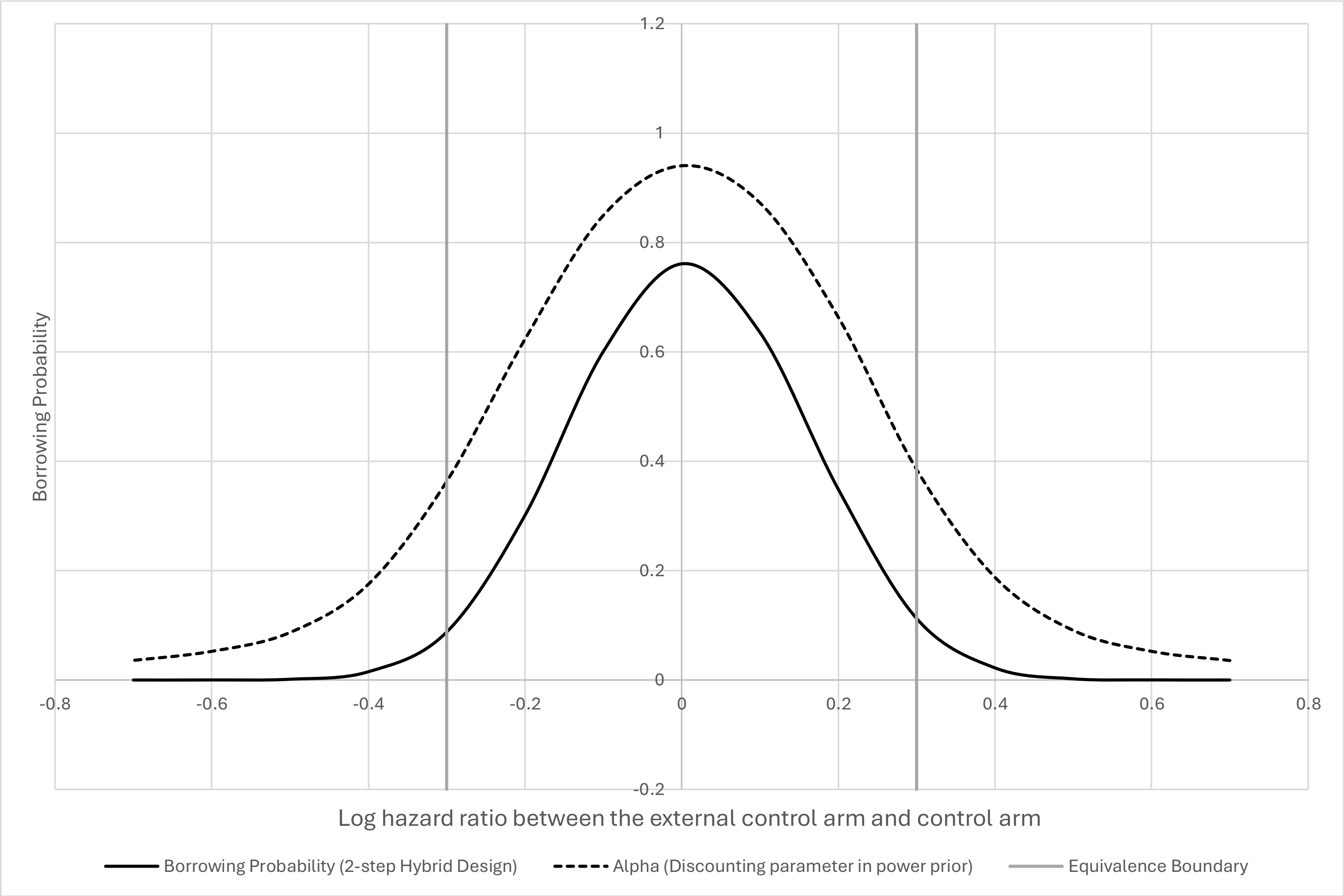

**Figure 4.** Bias under different log HR between control arm and external control arm for 2-step hybrid design and power prior

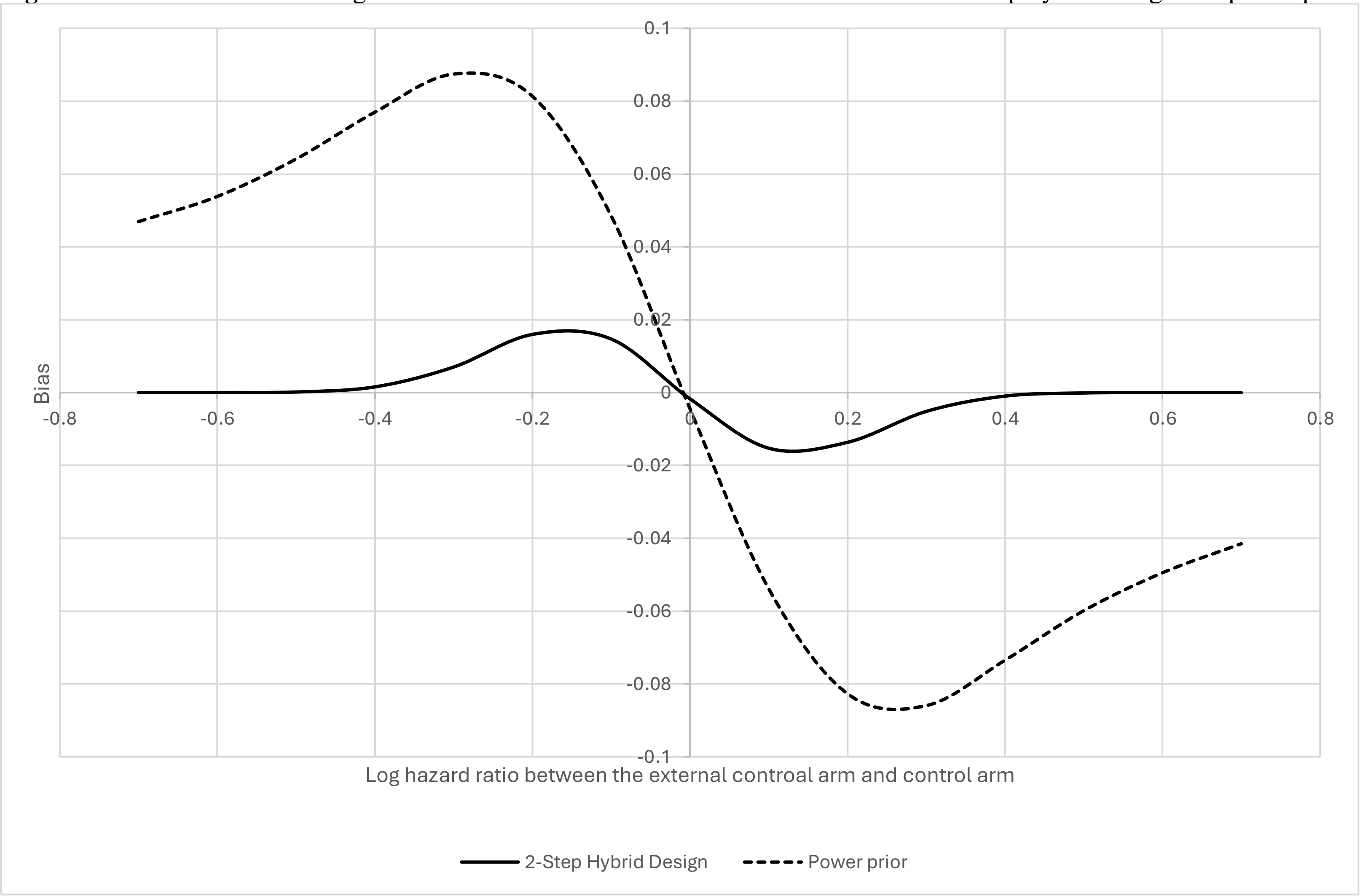

**Figure 5.** Simulation results under the null hypothesis: Type I error (in percentages) under different log HR between control arm and external control arm

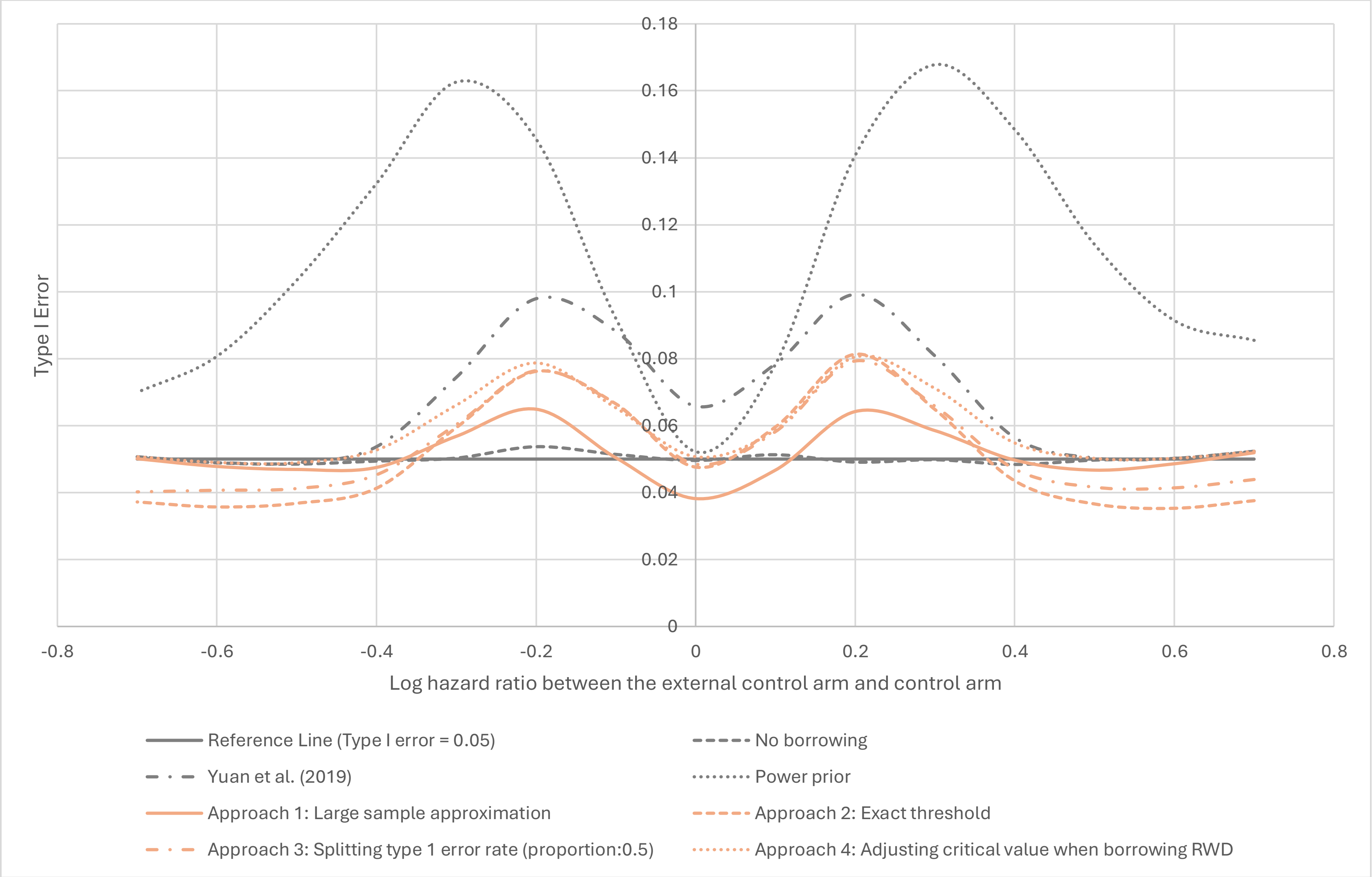

**Figure 6.** Simulation results under the alternative hypothesis: Power (in percentages) under different log HR between control arm and external control arm

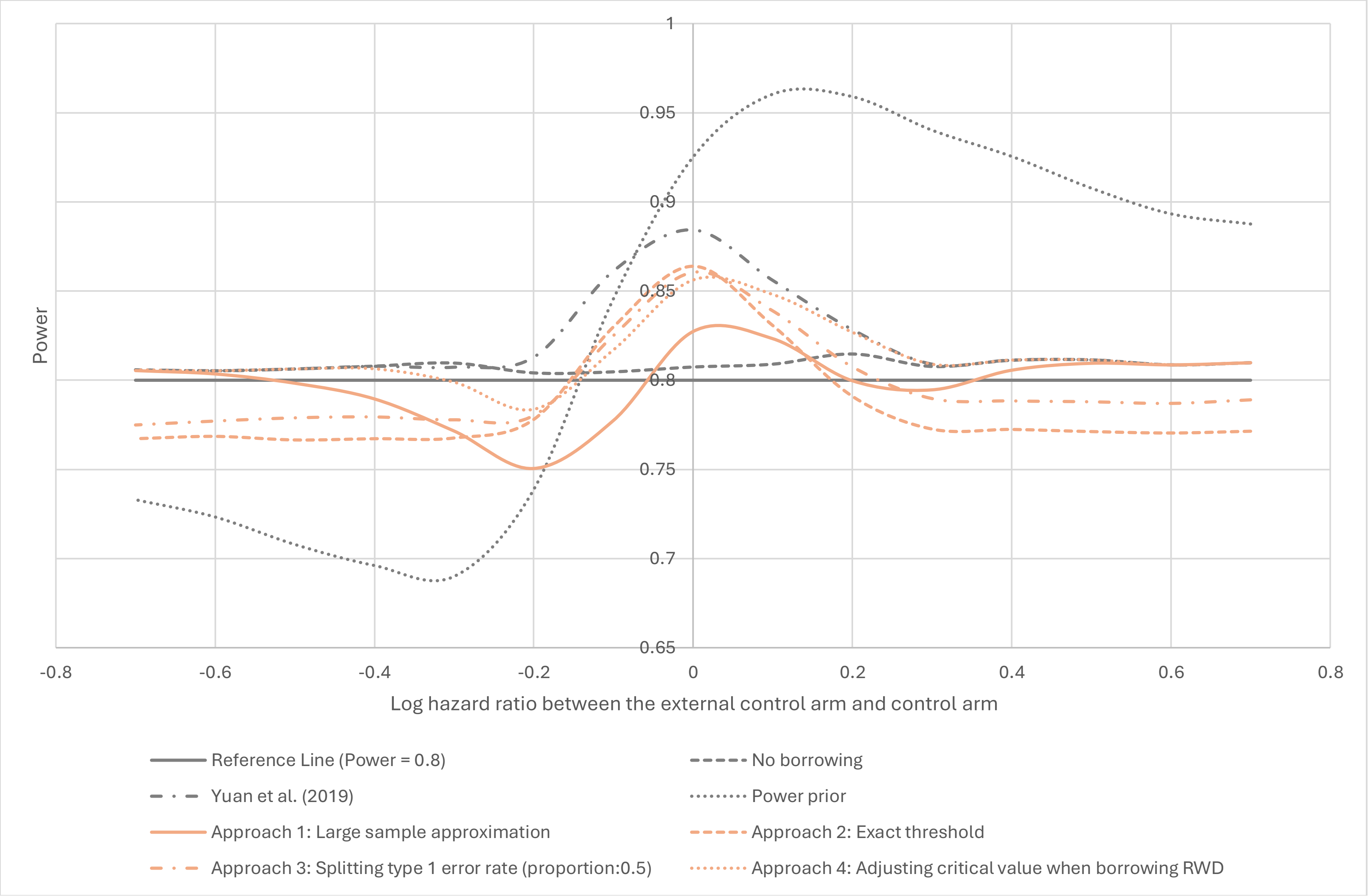